# An improved physics model for multi-material identification in photon counting CT

Xu Dong[a], Olga V. Pen[a], Zhicheng Zhang[a,b], Guohua Cao[a*]
[a]Virginia Tech-Wake Forest University School of Biomedical Engineering and Sciences, Virginia Polytechnic Institute and State University, Blacksburg, VA, USA, 24061; [b]Institue of Biomedical and Health Engineering, Shenzhen Institutes of Advanced Technology, Chinese Academy of Sciences, Shenzhen, Guangdong, China, 518055

**ABSTRACT**

Photon-counting computed tomography (PCCT) with energy discrimination capabilities hold great potentials to improve the limitations of the conventional CT, including better signal-to-noise ratio (SNR), improved contrast-to-noise ratio (CNR), lower radiation dose, and most importantly, simultaneous multiple material identification. One potential way of material identification is via calculation of effective atomic number ($Z_{eff}$) and effective electron density ($\rho_{e_{eff}}$) from PCCT image data. However, the current methods for calculating effective atomic number and effective electron density from PCCT image data are mostly based on semi-empirical models and accordingly are not sufficiently accurate. Here, we present a physics-based model to calculate the effective atomic number and effective electron density of various matters, including single element substances, molecular compounds, and multi-material mixtures as well. The model was validated over several materials under various combinations of energy bins. A PCCT system was simulated to generate the PCCT image data, and the proposed model was applied to the PCCT image data. Our model yielded a relative standard deviations for effective atomic numbers and effective electron densities at less than 1%. Our results further showed that five different materials can be simultaneously identified and well separated in a $Z_{eff} - \rho_{e_{eff}}$ map. The model could serve as a basis for simultaneous material identification from PCCT.

## 1. INTRODUCTION

Photon-counting Computed Tomography (PCCT) has a potential to overcome the limitations of conventional CT such as better signal-to-noise ratio (SNR), improved contrast-to-noise ratio (CNR), lower radiation dose, and most importantly, simultaneous multiple material identification [1]. Different from conventional CT which is energy integrative, PCCT is energy discriminative and thus can generate multiple CT image datasets corresponding to the multiple preset energy bins. Therefore, the richness of the image information from PCCT opens the possibility of achieving the multiple material identification capability for CT.

Previously, several methods have been proposed to identify or separate materials based on PCCT data. Butzer *et. al.* used principal component analysis method to extract the difference between materials [2]. Wang *et al.* proposed an angular separation method in the attenuation map to separate materials [3]. Another method is to calculate effective atomic number and effective electron density for the imaged materials [4, 5]. Because effective atomic number and effective electron density can be at least *nominally* considered intrinsic properties of materials, characterizing a material by its effective atomic number and effective electron density is potentially more accurate. The goal of this study is to develop a physics model to calculate the effective atomic number and effective electron density from PCCT data to achieve multiple material identification from PCCT.

Calculating effective atomic number for compounds and mixtures from CT image data has been a classical problem in the CT field [6-15]. Interestingly, several different definitions for effective atomic number have been proposed. For example, effective atomic number was typically calculated with atomic percentage weighting from the formula of a multi-element material [6, 14]. Others proposed to use the power law to approximate effective atomic number [8, 10], or calculate effective atomic number based on interpolation [11]. More comprehensive reviews about different definitions for effective atomic number can be found in the reference [12, 15].

One major problem for applying those definitions of effective atomic number into real-world applications is that those definitions are based on rather simple approximations, for example the power law approximation. Those simple approximations can lead to large error for some materials, which limits their use to general applications such as the various tissues/materials encountered in clinical imaging. Furthermore, those methods not always give consistent results. For example, some methods lead to energy-invariant effective atomic numbers, while others lead to energy-variant effective atomic numbers. One aim in this work is to address such inconsistency in the calculation of effective atomic numbers from PCCT data.

Hawkes and Jackson parametrized the x-ray attenuation coefficients based on the x-ray matter interaction physics, and derived exact formulas for effective atomic number and effective electron density for heterogeneous materials in 1980s [16, 17]. Because their model started from the basic physics *ab initio*, their formulas for effective atomic number and effective electron density can be valid for a broad range of scenarios. However, their formulas are very complicated and hard to compute, especially for compounds and mixtures. The details of Hawkes and Jackson's model will be explained more as follows.

In this work, our objective is to build a model to calculate effective atomic number and effective electron density so that we can rely on them for multiple material identification in biomedical applications. To achieve this purpose, a balance between the physical accuracy and the numerical computability of the model was made. Our model is developed based on the x-ray matter interaction physics [16] [17], so that the model can have high accuracy. We also made simplifications to make the model computable in calculation of effective atomic number and effective electron density for different materials from PCCT image data.

The paper is organized as follows. After the introduction, in section II we explained our model for calculating effective atomic number and effective electron density, and validated the accuracy of the model over some known reference materials. We then carried out PCCT simulations and used the data from simulations to demonstrate the feasibility of the model in calculating effective atomic number and effective electron density from PCCT imaging data. The results are shown in section III, and the discussions and conclusions are in section IV.

## 2. METHODS

### 2.1 Overview of the model

Linear attenuation coefficient ($\mu$) of a material is determined by the photon energy ($E$) and the intrinsic properties of the material. For single elements, the intrinsic properties are atomic number ($Z$) and electron density ($\rho_e$). According to the Hawkes and Jackson [16, 17], the linear attenuation coefficient ($\mu$) of a single element can be parameterized by the following equation:

$$\mu = f(E, Z, \rho_e) = \rho_e(Z^4 F(E, Z) + G(E, Z)) \qquad (1)$$

where $Z^4 F(E, Z)$ corresponds to the photoelectric cross-section, and $G(E, Z)$ corresponds to the scattering cross-section term. The detailed derivations of the equations $F(E, Z)$ and $G(E, Z)$ can be found in [16, 17]. We also summarized the derivations in APPENDIX in this paper.

Compounds and mixtures are composited by a bunch of single elements with different atomic numbers ($Z_i$) and electron densities ($\rho_{e_i}$). For a compound or mixture composed of $n$ different elements, Hawkes and Jackson provides a detailed parameterization of $\mu$ [17]. The equation can be summarized as:

$$\mu = g(E, Z_1, Z_2, \ldots, Z_n, \rho_{e_1}, \rho_{e_2}, \ldots, \rho_{e_n}) \qquad (2)$$

This equation (2) is based on the elemental composition of a specific material, which has $2n + 1$ parameters and makes the formula unpractical to be used with PCCT data.

Inspired by equation (1) for the parameterization of $\mu$ for single elements, here we define two corresponding parameters for compounds and mixtures: the effective atomic number ($Z_{eff}$) and the effective electron density ($\rho_{e_{eff}}$)[1], and we expect the linear attenuation coefficient ($\mu$) for compounds and mixtures can be parameterized in a similar way as for single elements:

$$\mu = f\left(E, Z_{eff}, \rho_{e_{eff}}\right) = \rho_{e_{eff}}(Z_{eff}^4 F(E, Z_{eff}) + G(E, Z_{eff})) \quad (3)$$

With this definition for effective atomic number and effective electron density, equation (3) can be applicable to any material, be it single element, compound, or mixture. Then, $Z_{eff}$ and $\rho_{e_{eff}}$ can be iteratively computed from several measurements of $E$ and $\mu$ according to equation (3). The iterative algorithm we chose in the study was Levenberg–Marquardt algorithm. Therefore, using the definition of effective atomic number and effective electron density from equation (3) and the Levenberg–Marquardt algorithm, one can calculate the energy-invariant $Z_{eff}$ and $\rho_{e_{eff}}$ from several groups of $(E, \mu)$ measurements.

## 2.2 Validation of the model

To validate the model, we manually tested the accuracy of the calculated $Z_{eff}$ and $\rho_{e_{eff}}$ under various x-ray energy conditions for a broad range of reference materials. The materials contain single-element substance (carbon, sodium, aluminum, and calcium), compounds (acetone, water, silicon dioxide, sodium chloride, and calcium peroxide), and mixtures (70% ethanol solution in water (v/v), 0.9% sodium chloride solution in water (m/v), and 10% sodium chloride solution in water (m/v).

For each material, we randomly chose photon energy ($E$) in the range from 30 keV to 120 keV and obtained the corresponding linear attenuation coefficients ($\mu$) from the NIST database [18]. Then we repeated the calculation 10,000 times with different randomly chosen photon energies to test the robustness of the model. Furthermore, since $Z_{eff}$ and $\rho_{e_{eff}}$ are iteratively calculated, the number of $(E, \mu)$ combination as the input of the model is flexible, which can be any number more than one. Therefore, we tested the number of $(E, \mu)$ combination from 2 to 8, to validate its accuracy over different input scenarios. From the 10,000 calculations, we calculated the mean value and the standard deviation of $Z_{eff}$ and $\rho_{e_{eff}}$. The relative standard deviation (defined as $\frac{mean}{standard\ deviation}$) was used to measure the accuracy of the model.

## 2.3 Numerical simulation of PCCT

To investigate the feasibility of applying the model for PCCT data, we conducted numerical simulation of a PCCT system to collect PCCT image data and calculate $Z_{eff}$ and $\rho_{e_{eff}}$ from them.

The simulated fan-beam PCCT system is illustrated in Fig. 1. The PCCT system consisted of an x-ray source, a filter (2.10 mm Aluminum), a phantom, and a photon counting detector. The geometric parameters are listed in Table 1. The excitation spectrum used in the simulation is generated from SpekCalc program [19].

Two phantoms are used in the simulation. One is a water phantom (H2O). Another is a contrast phantom which consists of water as the background material and four uniformly distributed slots around the center. Acetone (C3H6O), silicon dioxide (SiO2), sodium chloride (NaCl), and calcium peroxide (CaO2) were inserted into each slot respectively. The illustration of the simulated phantoms is shown in Fig. 2.

The PCCT numerical simulation mainly consists of two parts: the forward projection model and the photon counting detector model. The line-based Siddon's algorithm [20] was used to model the CT forward projection. After the photons reach the detector, all the incident photons will go through the photon counting detector model. In reality, a real photon counting detector has many spectrum degradation effects. To model the complexity of a real detector, we adapted the photon counting detector model developed by Taguchi et al. [21] to simulate the realistic detector response, including photon noise and detector noise, charge sharing, K-escape x-ray, etc. Meanwhile, for the purpose of comparison, an ideal detector model was used in the simulation, which simply assumes the x-ray spectrum outputted by the detector is same as the spectrum incident onto the detector surface without any distortion in the detector response.

---

[1] We use the name of effective electron density here instead of electron density, because this value of the effective electron density is calculated from equation (3). It may or may not be exactly the real electron density of the material itself. The purpose of defining such parameter is to use it to characterize materials, which could or could not be the real electron density for the material. We are using the terminology of effective electron density here in order to avoid the potential conflict with the concept of real electron density of a material.

After collecting the projection data, Filtered Backpropagation (FBP) algorithm was used for image reconstruction to generate PCCT images.

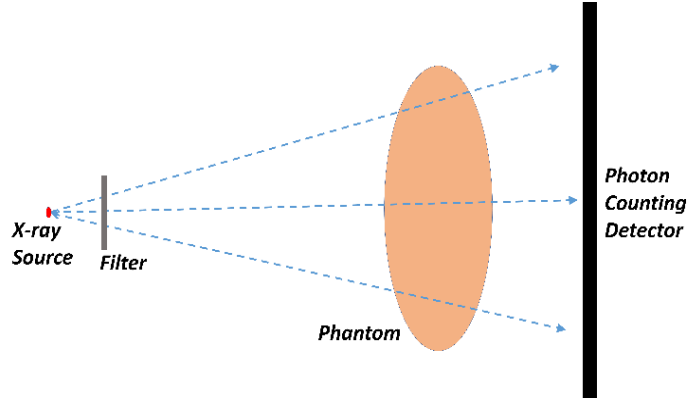

Figure 1.   Schematics of the simulated PCCT system.

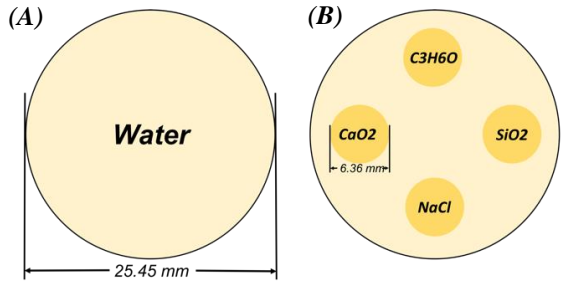

Figure 2.   Schematics of the phantoms used in the simulation. (A) The water phantom. (B) The contrast phantom that is made by water as the background material and with four slots around the center each filled with C3H6O, SiO2, NaCl, and CaO2, respectively.

Table 1. Parameters of the PCCT system.

| | |
|---|---|
| Source voltage | 120 kV |
| Filter | 2.10 mm Aluminum |
| Source to isocenter distance | 147 mm |
| Source to detector distance | 515 mm |
| Detector pixel size | 0.5 mm |
| Detector resolution | 256 pixels |
| Data acquisition | 1 projection per degree, 360 projections in total |
| Photon statistics | 1.0e6 photons per ray |

## 2.4 Calculation of $Z_{eff}$ and $\rho_{e_{eff}}$ from PCCT data

In the PCCT simulation, eight energy bins were pre-set from 50 keV to 120 keV with a fixed energy bin width of 10 keV, as listed in Table 2. Therefore, eight reconstructed images were generated in one PCCT scan, with each image corresponding to one energy bin. We first calculated the effective energy ($E_{eff}$) for each energy bin defined as the average of the x-ray photon energies weighted by the x-ray photon intensities at each energies:

$$E_{eff} = \frac{\sum_{E_{low}}^{E_{high}} E_i * I_i}{\sum_{E_{low}}^{E_{high}} I_i} \qquad (4)$$

where $E_{low}$ and $E_{high}$ are the boundaries of the energy bin, $E_i$ is the energy of each x-ray photon in the energy bin, and $I_i$ is the intensity of the x-rays with the energy of $E_i$. By equation (4), we can use $E_{eff}$ to represent the effective energy of interacting x-ray photons in each energy bin.

After obtaining $E_{eff}$ for every energy bin, the corresponding linear attenuation coefficients ($\mu$) corresponding to the energy bin can be obtained from the pixel values in the CT image. To eliminate the interference of the image noise, we calculated the averaged linear attenuation coefficient ($\mu$) for each material by averaging the image pixel values in the Region of Interest (ROI) illustrated in Fig. 3. As a result, a combination of the interacting x-ray photon energy ($E$) and the linear attenuation coefficient ($\mu$) can be obtained for each material at each energy bin. Eight combinations of $(E, \mu)$ can be obtained from the PCCT image data sets with eight energy bins.

Then we fed the eight $(E, \mu)$ combinations into the model to calculate $Z_{eff}$ and $\rho_{e_{eff}}$. To compare the accuracy, the $Z_{eff}$ and $\rho_{e_{eff}}$ were calculated from PCCT data by the realistic detector model, PCCT data by the ideal detector model, as well as the linear attenuation coefficient values taken from the NIST database [18]. The calculated $Z_{eff}$ and $\rho_{e_{eff}}$ from NIST database data were treated as the ground truth for comparison. A $Z_{eff} - \rho_{e_{eff}}$ plot was generated to display the result as shown in Fig. 4. We also computed the relative errors of the calculated $Z_{eff}$ and $\rho_{e_{eff}}$ from the PCCT data by the realistic detector and the ideal detector in comparison to the ground truth, which is shown in Table 4.

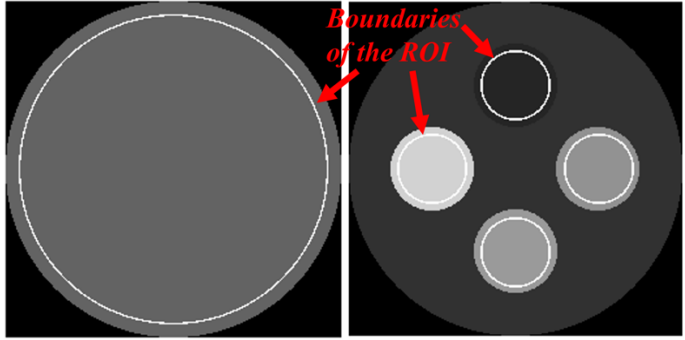

Figure 3. The illustration of the ROIs in the reconstructed images to calculate the averaged linear attenuation coefficients ($\mu$). The left image is one reconstructed image of the water phantom; the right image is one reconstructed image of the contrast phantom.

Table 2. The setup of the energy bins

| Energy Bin | Energy Range | Corresponding Effective Energy |
| --- | --- | --- |
| Bin 1: | 50 – 60 keV | 56.19 keV |
| Bin 2: | 60 – 70 keV | 65.23 keV |
| Bin 3: | 70 – 80 keV | 74.84 keV |
| Bin 4: | 80 – 90 keV | 84.79 keV |
| Bin 5: | 90 – 100 keV | 94.71 keV |
| Bin 6: | 100 – 110 keV | 104.53 keV |
| Bin 7: | 110 – 120 keV | 113.38 keV |

## 3. RESULT

### 3.1 Validation of the model

The relative standard deviations of the computed $Z_{eff}$ and $\rho_{e_{eff}}$ are reported in Table 3. The result shows that the relative standard deviations for all the materials are mostly less than 1% (161 out of 168). The small standard deviations indicate that the $Z_{eff}$ and $\rho_{e_{eff}}$ calculated from the model are not dependent on the input energies. Furthermore, as the number of $(E, \mu)$ combinations increases, the relative standard deviation becomes smaller. This trend is valid for all the studied materials, which implies that the model has higher accuracy with more energy bins as inputs. For instance, when using eight $(E, \mu)$ combinations as the input, the relative standard deviations for all the studied materials are less than 0.38%.

Table 3. The relative standard deviations of the calculated effective atomic numbers ($Z_{eff}$) and effective electron densities ($\rho_{eff}$) under different numbers of (μ, E) combinations and repeated with 10,000 different energies.

| Number of (μ, E) combinations | Carbon (C) | | Sodium (Na) | | Aluminum (Al) | | Calcium (Ca) | | Acetone (C3H6O) | |
|---|---|---|---|---|---|---|---|---|---|---|
| | Relative Standard Deviation of $Z_{eff}$ | Relative Standard Deviation of $\rho_{eff}$ | Relative Standard Deviation of $Z_{eff}$ | Relative Standard Deviation of $\rho_{eff}$ | Relative Standard Deviation of $Z_{eff}$ | Relative Standard Deviation of $\rho_{eff}$ | Relative Standard Deviation of $Z_{eff}$ | Relative Standard Deviation of $\rho_{eff}$ | Relative Standard Deviation of $Z_{eff}$ | Relative Standard Deviation of $\rho_{eff}$ |
| 2 | 5.45% | 0.22% | 0.63% | 0.19% | 5.45% | 0.22% | 0.37% | 0.18% | 0.16% | 1.45% |
| 3 | 0.84% | 0.05% | 0.15% | 0.05% | 0.84% | 0.05% | 0.09% | 0.05% | 0.09% | 0.75% |
| 4 | 0.43% | 0.03% | 0.07% | 0.03% | 0.43% | 0.03% | 0.05% | 0.03% | 0.07% | 0.54% |
| 5 | 0.27% | 0.02% | 0.05% | 0.02% | 0.27% | 0.02% | 0.03% | 0.02% | 0.06% | 0.44% |
| 6 | 0.20% | 0.02% | 0.04% | 0.02% | 0.20% | 0.02% | 0.02% | 0.02% | 0.05% | 0.36% |
| 7 | 0.16% | 0.02% | 0.03% | 0.02% | 0.16% | 0.02% | 0.02% | 0.02% | 0.04% | 0.31% |
| 8 | 0.14% | 0.01% | 0.03% | 0.02% | 0.14% | 0.01% | 0.02% | 0.02% | 0.04% | 0.27% |

| Number of (μ, E) combinations | Water (H2O) | | Silicon Dioxide (SiO2) | | Sodium Chloride (NaCl) | | Calcium Peroxide (CaO2) | | 70% Ethanol | |
|---|---|---|---|---|---|---|---|---|---|---|
| | Relative Standard Deviation of $Z_{eff}$ | Relative Standard Deviation of $\rho_{eff}$ | Relative Standard Deviation of $Z_{eff}$ | Relative Standard Deviation of $\rho_{eff}$ | Relative Standard Deviation of $Z_{eff}$ | Relative Standard Deviation of $\rho_{eff}$ | Relative Standard Deviation of $Z_{eff}$ | Relative Standard Deviation of $\rho_{eff}$ | Relative Standard Deviation of $Z_{eff}$ | Relative Standard Deviation of $\rho_{eff}$ |
| 2 | 0.63% | 0.19% | 0.48% | 0.20% | 0.28% | 0.23% | 0.29% | 0.39% | 7.48% | 0.99% |
| 3 | 0.15% | 0.05% | 0.13% | 0.05% | 0.09% | 0.09% | 0.17% | 0.23% | 3.73% | 0.54% |
| 4 | 0.07% | 0.03% | 0.07% | 0.04% | 0.06% | 0.07% | 0.14% | 0.19% | 2.02% | 0.35% |
| 5 | 0.05% | 0.02% | 0.05% | 0.03% | 0.05% | 0.06% | 0.12% | 0.17% | 1.12% | 0.23% |
| 6 | 0.04% | 0.02% | 0.04% | 0.02% | 0.04% | 0.05% | 0.11% | 0.15% | 0.82% | 0.19% |
| 7 | 0.03% | 0.02% | 0.03% | 0.02% | 0.04% | 0.04% | 0.10% | 0.14% | 0.46% | 0.17% |
| 8 | 0.03% | 0.02% | 0.03% | 0.02% | 0.03% | 0.04% | 0.09% | 0.13% | 0.38% | 0.15% |

| Number of (μ, E) combinations | Saline Solution | | 10% NaCl | |
|---|---|---|---|---|
| | Relative Standard Deviation of $Z_{eff}$ | Relative Standard Deviation of $\rho_{eff}$ | Relative Standard Deviation of $Z_{eff}$ | Relative Standard Deviation of $\rho_{eff}$ |
| 2 | 2.19% | 0.20% | 0.96% | 0.19% |
| 3 | 0.57% | 0.07% | 0.33% | 0.08% |
| 4 | 0.33% | 0.05% | 0.23% | 0.06% |
| 5 | 0.22% | 0.04% | 0.19% | 0.05% |
| 6 | 0.18% | 0.03% | 0.16% | 0.04% |
| 7 | 0.16% | 0.03% | 0.14% | 0.04% |
| 8 | 0.15% | 0.03% | 0.12% | 0.03% |

### 3.2 The linear attenuation coefficients in PCCT image data sets

As shown in Fig. 4, we can see all the five materials can be clearly separated in the $Z_{eff} - \rho_{e_{eff}}$ map, and the calculated $Z_{eff}$ and $\rho_{e_{eff}}$ from the PCCT data (both realistic detector and ideal detector) are well correlated with the calculated $Z_{eff}$ and $\rho_{e_{eff}}$ from the NIST data (ground truth). We can clearly separate and identify each material in the $Z_{eff} - \rho_{e_{eff}}$ map by comparing the calculated $Z_{eff}$ and $\rho_{e_{eff}}$.

We also computed the relative errors of the calculated $Z_{eff}$ and $\rho_{e_{eff}}$ from the PCCT data by comparing them with the $Z_{eff}$ and $\rho_{e_{eff}}$ calculated from the NIST data. As listed in Table 4, the absolute relative errors of the calculated $Z_{eff}$ across all the five materials are averaged to be 0.78% for the ideal detector model and 3.03% for the realistic detector model. And the absolute relative errors of the calculated $\rho_{e_{eff}}$ across all the five materials are averaged to be 0.81% for the ideal detector model and 1.39% for the realistic detector model.

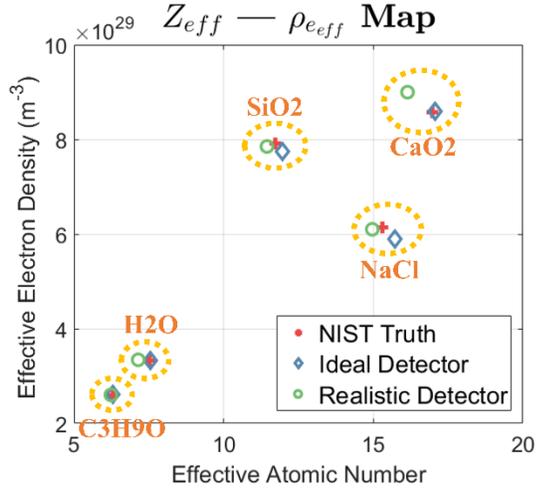

Figure 4. The calculated effective atomic number ($Z_{eff}$) and effective electron densities ($\rho_{e_{eff}}$) for the five materials (acetone, water, silicon dioxide, sodium chloride, and calcium peroxide) from different data sources. The red dots represent the results calculated based on the NIST database, which are taken as the ground truth; the blue dots represent the results calculated from the simulated PCCT data by the ideal detector model; the green dots represent the results calculated from the simulated PCCT data by the realistic detector model.

Table 4.(A) The relative errors of the calculated effective atomic numbers ($Z_{eff}$) from different detector models

|  | C3H6O | H2O | SiO2 | NaCl | CaO2 |
|---|---|---|---|---|---|
| Ideal Detector Model | 0.34% | 0.39% | 1.10% | 1.61% | 0.46% |
| Realistic Detector Model | -0.84% | -4.84% | -2.30% | -2.28% | -4.90% |

The ground truth are the effective atomic numbers that are calculated from NIST database data.

Table 4.(B) The relative errors of the calculated effective electron densities ($\rho_{e_{eff}}$) from different detector models

|  | C3H6O | H2O | SiO2 | NaCl | CaO2 |
|---|---|---|---|---|---|
| Ideal Detector Model | 0.68% | 0.10% | -1.12% | -2.12% | 0.05% |
| Realistic Detector Model | 0.22% | 0.41% | -0.72% | -0.78% | 4.82% |

The ground truth are the effective electron densities that are calculated from the NIST database data.

## 4. DISCUSSION AND CONCLUSION

In this study, we developed a model to calculate effective atomic number and effective electron density based on x-ray interaction physics. We validated the accuracy of the model for many different materials. Furthermore, we incorporated the model with PCCT data and demonstrated the feasibility of the model to achieve material separation and identification from PCCT image data.

As the validation of the model, we calculated $Z_{eff}$ and $\rho_{e_{eff}}$ under various energy conditions for many materials. The result shows that, given a specific material, the calculated $Z_{eff}$ and $\rho_{e_{eff}}$ from the model demonstrate good accuracy and robustness to different energy conditions (as shown in Table 3, most of the relative standard deviations are within 1%). The good accuracy of the model indicates that the model can be used to characterize different materials in an accurate and reliable manner.

To investigate the feasibility of incorporating the model with PCCT data, we conducted PCCT system simulation to generate PCCT image data. In the simulation, we adopted a realistic detector model [21] that includes many detector degradation effects (noise, charge sharing, K-escape x-ray, *et al.*), and also simulated an ideal detector which assumes perfect detector response for comparison. Then we calculated the $Z_{eff}$ and $\rho_{e_{eff}}$ of the five simulated materials from PCCT data using the model. As can be seen in the $Z_{eff} - \rho_{e\,eff}$ map (Fig. 4), the five materials are well separated and can be easily identified in the map. This confirms the usability of the model to achieve multi-material identification from PCCT.

The idea of employing effective atomic number to characterize materials has been explored by some previous studies [22-24], however they have some limitations. Firstly, one model developed to calculate effective atomic number is adapted from dual-energy CT [25, 26], which can lead to different effective atomic number values under different energy conditions. Secondly, the accuracy of this model varies largely for different materials. For example, the relative errors of calculated effective atomic number can be as small as 0.5% for Carbon and as large as 16.8% for Titanium as reported in [22]. Here we reported an improved model to calculate effective atomic number that is energy-invariant. The model retains good accuracy for a broad range of materials. When applying the model with PCCT image data, the biggest error is 4.9%. An improved accuracy is achieved from our model.

The model we developed in the study has some limitations. Firstly, The PCCT data from the realistic detector model leads to larger errors in $Z_{eff}$ and $\rho_{e_{eff}}$ compared to PCCT data from the ideal detector model. This is because the realistic detector data are degraded by many detector defects. As the detector manufacturing technique improves in the future, we expect our model could become suitable for achieving material identification from PCCT via effective atomic number and effective electron density. Secondly, the model has poor accuracy for high Z materials. As can be observed in Table 3, as the effective atomic number becomes higher, the relative error becomes higher as well. In our study, we calculated $Z_{eff}$ for Iodine using our model, and the resulted $Z_{eff}$ is 7% off from the real atomic number of iodine. We think the discrepancy is mainly because the model does not handle the K-edge absorption well. As we know, K-edge absorption plays a significant role in attenuation coefficients for high Z materials. On the other hand, our model demonstrates very good accuracy for materials with effective atomic number as high as 20 (Calcium), as shown in Table 3. Since the biological materials mostly have effective atomic number lower than 20, the model can still be applied to biological imaging applications with relatively good accuracy.

In summary, we developed a model to calculate effective atomic number and effective electron density to characterize materials from PCCT image data. The feasibility of applying the model to PCCT system was demonstrated by the simulated PCCT data. Different materials were clearly identified by the calculated effective atomic numbers and effective electron densities. The model could serve as a basis for simultaneous material identification from PCCT.

## 5. APPENDIX

When an x-ray beam passes through an object, the intensity of the x-ray beam would be weakened due to the attenuation by the object. And the interaction between the x-rays and the imaged object is characterized by the Beer-Lambert law:

$$\frac{I}{I_0} = \exp(-\mu x) \qquad (5)$$

where $I$ is the intensity of the x-ray beam after the attenuation, $I_0$ is the intensity of the x-ray beam before the attenuation, and $\mu$ is the linear attenuation coefficient, which is determined jointly by the x-ray photon energy ($E$) and the intrinsic properties of the interacting material itself.

For single elements, the intrinsic properties of the material can be summarized as the atomic number ($Z$) and the electron density ($\rho_e$). According to the Hawkes and Jackson [16, 17], the linear attenuation coefficient ($\mu$) of single elements can be parameterized by the following equation:

$$\mu = f(E, Z, \rho_e) = \rho_e(Z^4 F(E, Z) + G(E, Z)) \qquad (6)$$

where $Z^4 F(E, Z)$ corresponds to the photoelectric cross-section, and $G(E, Z)$ corresponds to the scattering cross-section term, which includes both the coherent scattering and the incoherent scattering.

Hawkes and Jackson derived the equations of $F(E, Z)$ and $G(E, Z)$ from the scratch on the basis of the physics of the interaction between an x-ray photon and an electron [16, 17]. The final expression $F(E, Z)$ is summarized as:

$$F(E,Z) = 4 * \sqrt{2} * \left(\frac{e^2}{\hbar c}\right)^4 * (m_e c^2)^{\frac{7}{2}} * \frac{\sigma^T}{E^{\frac{7}{2}}} * S(E,Z) * N(Z) * (1 + \mathcal{F}(\beta)) \tag{7}$$

where $e$ is the electron charge, $\hbar$ is the Dirac constant, $c$ is the light speed, $m_e$ is the rest mass of an electron, $r_e$ is the classical electron radius, and $\sigma^T$ represents the Thomson cross-section, which equals to $\frac{8*\pi}{3}r_e^2$.

The function of $S(E,Z)$ is used to apply correction to the Born approximation from the Stobbe's study [27]:

$$S(E,Z) = 2\pi * \left(\frac{\varepsilon_K}{E-\varepsilon_K}\right)^{0.5} * \frac{\exp(-4*n_1*\cot^{-1} n1)}{1-\exp(-2\pi*n_1)} \tag{8}$$

where $\varepsilon_K = \frac{Z^2 m_e e^4}{2\hbar^2}$, and $n_1 = \left(\frac{\varepsilon_K}{E-\varepsilon_K}\right)^{0.5}$.

The term of $(1 + \mathcal{F}(\beta))$ is used to incorporate the relativistic effects [28], which is expressed as: $1 + \mathcal{F}(\beta) = 1 + 0.143\beta^2 + 1.667\beta^8$, where $\beta = \sqrt{\frac{2E}{m_e c^2}}$.

As for the $G(E,Z)$ function, the final expression is summarized as:

$$G(E,Z) = \sigma^{KN} + \frac{(1-Z^{b-1})}{Z} * \left(\frac{Z}{Z'}\right)^2 * \left(\frac{3}{8} * \sigma^T\right) * \int_{-1}^{1} (1 + \cos^2(\theta)) * [\mathbb{F}(x, Z')]^2 d(\cos(\theta)) \tag{9}$$

where $Z'$ is the atomic number of a standard element serving as the reference to calculate the coherent scattering cross-section, $E' = \left(\frac{Z}{Z'}\right)^{\frac{1}{3}} * E$, and $b$ is the empirically established parameter that varies with different ranges of $Z$. We used $Z' = 8$ and $b = 0.5$ in our implementation, as suggested in the Hubbell's study [29] for the modeling of the soft tissue like materials.

$\sigma^{KN}$ refers to the Klein-Nishina cross-section [30]. It is used to model the coherent scattering for the interaction between a photon and a free electron, the equation is expressed as:

$$\sigma^{KN} = 2\pi * r_e^2 * \left(\frac{1+k}{k^2} * \left(2 * \frac{1+k}{1+2k} - \frac{\ln(1+2k)}{k}\right) + \frac{\ln(1+2k)}{2k} - \frac{1+3k}{(1+2k)^2}\right) \tag{10}$$

where $k = \frac{E}{m_e c^2}$.

$\mathbb{F}(x, Z')$ is the atomic form factor function, where $x$ is momentum-transfer variable, which equals to $\frac{\sin\left(\frac{\theta}{2}\right)}{\lambda(A)}$, and $\lambda(A)$ is the photon wavelength in angstroms, which equals to $\frac{12.398520}{E'(keV)}$. The detailed explanation can be found in Hubbell's study [29]. And $\theta$ is the angle between the photon directions of travel prior to and following a scattering interaction.

As a summary, using the above parameterization of $\mu(E, Z, \rho_e)$, the relationship of the linear attenuation coefficient $\mu$, the x-ray photon energy $E$, and the atomic number ($Z$) and the electron density ($\rho_e$) for a single element is established.